\documentclass[amsmath,amssymb,showkeys,superscriptaddress,floatfix,prd,10pt,aps]{revtex4-2}
\usepackage{graphicx,epstopdf}
\usepackage[caption=false]{subfig}
\usepackage{multirow}
\usepackage{appendix}
\def\xslash{x\!\!\!\slash }

\def\vel{\left|}
\def\ver{\right|}
\usepackage{colortbl}
\usepackage{xcolor}
\usepackage[colorlinks=true, urlcolor=blue, linkcolor=green, citecolor=green]{hyperref}

\begin{document}

\title{Electromagnetic form factors of the $B_c$-like tetraquarks: molecular and diquark-antidiquark pictures}

\author{Ula\c{s} \"{O}zdem}%
\email[]{ulasozdem@aydin.edu.tr}
\affiliation{Health Services Vocational School of Higher Education, Istanbul Aydin University, Sefakoy-Kucukcekmece, 34295 Istanbul, T\"{u}rkiye}

\date{\today}
 
\begin{abstract}
In this study, we use the molecular and diquark-antidiquark tetraquark pictures to investigate magnetic and quadrupole moments of the $B_c$-like ground state tetraquarks with the QCD light-cone sum rules with quantum numbers $J^P = 1^+$. 
 In the numerical analysis, to obtain the magnetic and quadrupole moments of $B_c$-like tetraquark states molecular and diquark-antidiquark forms of interpolating currents, and photon distribution amplitudes have been used.   
The magnetic moments are acquired as  $ \mu_{Z_{{uc \bar u \bar b}}}^{Mol}=1.18^{+0.52}_{-0.40}~\mu_N$, $\mu_{Z_{{uc \bar u \bar b}}}^{Di}=3.05^{+1.19}_{-0.95}~\mu_N$, $\mu_{Z_{dc \bar d \bar b}}^{Mol}=0.32^{+0.18}_{-0.10}~\mu_N$,  and  $\mu_{Z_{dc \bar d \bar b}}^{Di}=2.38^{+0.95}_{-0.75}~\mu_N$. 
The hadrons' magnetic and quadrupole moments are another fundamental observable as their mass, which provides information on the underlying quark structure and dynamics. 
The results obtained in both pictures are quite different from each other. Any experimental measurement of the magnetic moments can provide an understanding of the internal structure of these states. We get nonzero but small values for the quadrupole moments of $B_c$-like tetraquark states showing non-spherical charge distributions.   Hopefully, the examinations given in this study will be helpful to an experimental search of them, which will be an interesting research subject. 
\end{abstract}

\maketitle

\section{motivation}\label{motivation}

The experimental discovery of exotic states began in 2003 with the observation of the X(3872)~\cite{Belle:2003nnu}, though they were suggested long ago that states other than standard hadrons could exist. 
Since 2003, scientists have been increasingly interested in investigating and understanding the nature of exotic states that differ from standard mesons and baryons. 
Analysis of exotic states plays a key role in understanding low-energy QCD and therefore it is quite important to search for them in experiments.  
So far, many exotic states have been observed by different experimental facilities.
Many different theoretical interpretations of these states have been suggested, such as tetraquarks, pentaquarks, hybrids, glueballs, and so on (see the reviews of Refs. ~\cite{Esposito:2014rxa,Esposito:2016noz,Olsen:2017bmm,Lebed:2016hpi,Nielsen:2009uh,Brambilla:2019esw,Agaev:2020zad,Chen:2016qju,Ali:2017jda,Guo:2017jvc,Liu:2019zoy,Yang:2020atz,Dong:2021juy,Dong:2021bvy,Meng:2022ozq, Chen:2022asf}, and references
therein, for further details).%
%For the recent experimental and theoretical progress on the exotic states see, for instance,  Refs. 

 Most of the discovered tetraquark states belong to the class of so-called hidden-charm or bottom states including the $c \bar c$ or $b \bar b$ pair. However, the first principles of QCD do not prohibit the existence of open-flavor tetraquarks.  The $B_c$-like tetraquarks belong to another type of exotic state. Although these states have not been observed experimentally, they have attracted the attention of physicists \cite{Zhang:2009vs,Zhang:2009em,Sun:2012sy,Albuquerque:2012rq,Chen:2013aba, Agaev:2016dsg,Agaev:2017uky,Wang:2020jgb,Wang:2019xzt,Wu:2018xdi,Ortega:2020uvc}. 
 In Refs.~\cite{Zhang:2009vs, Zhang:2009em}, Zhang et al., have employed to calculate the masses for $[Q \bar q][\bar Q^{(')}q]$ and $[Q \bar s][\bar  Q^{(')}s]$ molecular states within the QCD sum rules, including the contributions of the operators up to dimension six in OPE.   
In Ref.~\cite{Sun:2012sy}, Sun et al., have studied the interaction between the S-wave $D^{(*)}/D^{(*)}_s$ meson and S-wave $B^{(*)}/B^{(*)}_s$ meson in the one-boson exchange model and they predicted the existence of many $B_c$-like molecular states.  
 In Ref.~\cite{Albuquerque:2012rq}, Albuquerque et al., have studied the mass of the exotic $B_c$-like molecular states using QCD sum rules and they predicted for these states masses around $7.0$ GeV. 
 In Ref.~\cite{Chen:2013aba}, Chen et al., have performed QCD sum rule analysis and extracted the masses of $[bc \bar q \bar q]$, $[bc \bar s \bar s]$ and $[qc \bar q \bar b]$ and $[sc \bar s \bar b]$ tetraquark states and they estimated that the tetraquark states $[qc \bar q \bar b]$ and $sc \bar s \bar b]$ lie below the thresholds of $D^{(*)}B^{(*)}$ and $D^{(*)}_s B^{(*)}_s$ respectively. 
 In Refs.~\cite{Agaev:2016dsg, Agaev:2017uky},  Agaev et al.,  have studied the spectroscopic parameters for $B_c$-like tetraquark states $[c q][\bar b \bar q]$ and $[cs][\bar b \bar s]$ with quantum numbers $J^P=0^+,1^+$ through QCD sum rule technique and their analyzes indicated that the masses are about $6.97-7.06$ GeV for the $[c q][\bar b \bar q]$  tetraquark states and $7.01-7.30$ GeV for the $[cs][\bar b \bar s]$ tetraquark states. In addition to the mass calculations of these states, some strong decay channels were also investigated in these studies and the relevant coupling constants were obtained. 
 In Ref.~\cite{Wang:2020jgb}, Wang et al., studied the masses for fully open-flavor tetraquark states $[b c \bar q \bar s]$ and $[s c\bar q \bar b]$ with quantum numbers $J^P=0^+,1^+$  in the framework of QCD sum rule method and their computations showed that the masses are about $7.1-7.2$ GeV for the $[b c \bar q \bar s]$ tetraquark states and $7.0-7.1$ GeV for the $[s c\bar q \bar b]$ tetraquarks. 
 In Ref.~\cite{Wang:2019xzt}, Z.-G. Wang has constructed the diquark-antidiquark type current operators to study the axial-vector $B_c$-like tetraquark states with the QCD sum rules with quantum numbers $J^P=1^{+-},1^{++}$ and predicted mass of these states given as $M_{Z_{b \bar c}(1^{+-})}=7.30 \pm 0.08$ GeV and $M_{Z_{b \bar c}(1^{++})}=7.31 \pm 0.08$ GeV, respectively. 
 In Ref.~\cite{Wu:2018xdi}, Wu et al., calculated the spectra of the possible $Q_1q_2 \bar Q_3 \bar q_4~ (Q = b, c$ and  $q = n, s$ with $n = u, d)$ tetraquark states by using the chromomagnetic interaction model in the diquark-antiquark picture. 
 In Ref.~\cite{Ortega:2020uvc}, Ortega et al., have obtained the masses of $DB$, $DB^*$, $D^*B$, and $D^*B^*$ states through the quark model and their computations indicated that the masses are about $7.1-7.3$ GeV for the $B_c$-like tetraquark states.

 The studying these $B_c$-like systems: contrary to $c \bar c$ and $b\bar b$, the $B_c$-like tetraquarks cannot annihilate into gluons and therefore these states are very stable, with narrow widths. Because of these properties, they are quite valuable to study heavy-quark dynamics and understanding the dynamics of the QCD at a deeper level.  To better understand the internal structures of these states, it is also important to study decay channels such as strong, radiative, and electromagnetic together with their spectroscopic parameters. Calculation of the electromagnetic properties of the particles allows us to obtain important data about the substructure of the particles under investigation.  
 Since in the past magnetic moments estimations from this sort of model have been well accomplished, to distinguish among the possible configurations, it seems to be helpful to investigate also the magnetic moments of multiquark states.
 Furthermore, the magnetic moments of the hadrons are important measurables like their masses, which have substantial knowledge about the underlying quark configurations, and can be used to distinguish the preferred quark configurations from different theoretical approximations and deepen our understanding of the underlying dynamics.
Inspired by the above reasons, in this work, we will study the magnetic and quadrupole moments for the $B_c$-like ($Z_{c\bar b}$ for short) ground state tetraquarks with the quantum numbers $J^P  = 1^+$ in the method of QCD light-cone sum rules (LCSR) which is the powerful quantitative tool to investigate features of hadrons. 
The key idea and the defining characteristic of the LCSR are that the short-distance operator product expansion is replaced by the light-cone expansion in operators of increasing twist~\cite{Chernyak:1990ag, Braun:1988qv, Balitsky:1989ry}.
Over the past few decades, LCSR has shown to be a very robust method for investigating non-perturbative hadron properties such as form factors, coupling constants, and magnetic moments associated with conventional and unconventional hadron states. 
The applications of the QCD sum rules and LCSR to some $c \bar c$ and $b \bar b$ tetraquark states can be seen in Refs.~\cite{Ozdem:2022kck,Xin:2022bzt, Agaev:2022iha, Wang:2022xja, Ozdem:2021hka,Agaev:2016dev,  Chen:2015ata, Xu:2020evn, Ozdem:2021yvo, Xu:2020qtg, Wang:2020iqt, Ozdem:2017exj,Wang:2018rfw, Yu:2017bsj, Ozdem:2017jqh,  Sundu:2018nxt, Agaev:2017lmc, Wang:2019veq, Wang:2019hnw, Agaev:2017tzv, Agaev:2017foq,     Wang:2016mmg, Chen:2016oma, Chen:2015fsa, Wang:2017dce}.

 The article is organized as follows. In Sec. \ref{formalism}, we briefly introduce our notations and apply the LCSR method to evaluate the magnetic and quadrupole moments of $Z_{c\bar b}$ tetraquark states as molecular and diquark-antidiquark structures. In Sec. \ref{numerical}, the numerical analysis and discussions for the magnetic and quadrupole moments of the $Z_{c\bar b}$ tetraquark states are presented. The obtained results are summarized and discussed in Sec. \ref{sum}. 
 The appendixes include explicit expressions of the correlator used in computations of the magnetic moments of the $Z_{c\bar b}$ tetraquark states and some details about calculations.

 %\begin{widetext}
 
\section{Magnetic and quadrupole Moments from LCSR}\label{formalism}
To evaluate the magnetic and quadrupole moments of the $Z_{c\bar b}$ ground state tetraquarks within the LCSR, we start with the following correlator 
\begin{align}
 \label{edmn01}
\Pi _{\mu \nu }(p,q)=i\int d^{4}xe^{ip\cdot x}\langle 0|\mathcal{T}\{J_{\mu}(x)
J_{\nu }^{\dagger }(0)\}|0\rangle_{\gamma}, 
\end{align}%
%where 
where  $q$ is the momentum of the photon, the $\gamma$ stands for the external background electromagnetic field and $J_{\mu}(x)$ is the interpolating currents of the $Z_{c\bar b}$ ground state tetraquarks with the quantum numbers $ J^{P} = 1^{+}$.  %The $B_c$ states can be interpolated by compact tetraquark states as 
The corresponding molecular and diquark-antidiquark interpolating currents are given by
\begin{align}
J_{\mu }^{Mol}(x) &= [\bar q_a (x) i\gamma_5 c_a (x)][\bar b_b (x) \gamma_\mu q_b (x)],\\
%%%%%%%%%%%%%%%%
J_{\mu }^{Di}(x)&=[q_{a}^{T}(x)C\gamma _{5}c_{b}(x)][ \overline{q}_{a}(x)\gamma _{\mu }C%
\overline{b}_{b}^{T}(x)]
%\nonumber\\
%&
+ [q_{a}^{T}(x)C\gamma _{5}c_{b}(x)] [\overline{q}_{b}(x)\gamma _{\mu }C\overline{b}%
_{a}^{T}(x)],  
\label{curr}
\end{align}
where $C$ is the charge conjugation operator and  $q(x)$ denotes one of the $u(x)$ or  $d(x)$  quarks.

%%%%%%%%%%%%%%%%%%%%%%%%%%%%%%%%%%%%

To get LCSR for the magnetic and quadrupole moments we follow standard instructions of the LCSR method and express the correlator $\Pi_{\mu\nu} (p,q)$ in connection with the physical parameters of the $Z_{c\bar b}$ tetraquark states, which results in obtaining  $\Pi^{Had}_{\mu\nu} (p,q)$. From another side, the same correlator should be acquired concerning the quark-gluon degrees of freedom $\Pi^{QCD}_{\mu\nu} (p,q)$. Matching the coefficients of various Lorentz structures from two different representations of the same correlator and performing double Borel transformations and continuum subtraction to remove the effects of the continuum and higher states, we get 
LCSR for the magnetic and quadrupole moments of the $Z_{c\bar b}$ tetraquark states. %As a final step, Borel transform as well as continuum subtraction supplied by the quark-hadron duality ansatz are carried out to remove the undesirable effects coming from the higher states and continuum.

 We will begin our analysis by calculating the hadronic representation of the correlator. To do this, the correlator is computed by its fulfillment with the intermediate $Z_{c\bar b}$ tetraquark states where $p^2> 0$, $(p+q)^2 > 0$. By applying the four-integral over $x$ we obtain
\begin{align}
\label{edmn04}
\Pi_{\mu\nu}^{Had} (p,q) &= {\frac{\langle 0 \mid J_\mu (x) \mid
Z_{c\bar b}(p) \rangle}{p^2 - m_{Z_{c\bar b}}^2}} \langle Z_{c\bar b} (p) \mid Z_{c\bar b} (p+q) \rangle_\gamma %\nonumber\\
%& \times
\frac{\langle Z_{c\bar b} (p+q) \mid J_{\nu }^{\dagger } (0) \mid 0 \rangle}{(p+q)^2 - m_{Z_{c\bar b}}^2} + \mbox{higher states},
\end{align}
%where higher states are shown by dots.   
The matrix element
$\langle 0 \mid J_\mu (x) \mid Z_{c\bar b}(p) \rangle$ is given as
\begin{align}
\label{edmn05}
\langle 0 \mid J_\mu (x) \mid Z_{c\bar b} (p) \rangle = \lambda_{Z_{c\bar b}} \varepsilon_\mu^\theta\,,
\end{align}
with $ \varepsilon_\mu^\theta\ $ and $\lambda_{Z_{c\bar b}}$ being the polarization vector and residue of the $Z_{c\bar b}$ tetraquark states, respectively. 

The matrix element $\langle Z_{c\bar b}(p) \mid Z_{c\bar b} (p+q)\rangle_\gamma $ can be written  in connection with  the Lorentz invariant form factors as follows~\cite{Brodsky:1992px}:

\begin{align}
\label{edmn06}
\langle Z_{c\bar b}(p,\varepsilon^\theta) \mid  Z_{c\bar b} (p+q,\varepsilon^{\delta})\rangle_\gamma &= - \varepsilon^\tau (\varepsilon^{\theta})^\alpha (\varepsilon^{\delta})^\beta 
%\nonumber\\
%&
\Big[ G_1(Q^2)~ (2p+q)_\tau ~g_{\alpha\beta}  
%\nonumber\\
%&
+ G_2(Q^2)~ ( g_{\tau\beta}~ q_\alpha -  g_{\tau\alpha}~ q_\beta)
\nonumber\\ 
&
- \frac{1}{2 m_{Z_{c\bar b}}^2} G_3(Q^2)~ (2p+q)_\tau 
%\nonumber\\
%&\times ~
q_\alpha q_\beta  \Big],
\end{align}
where the polarization vectors of the initial and final $Z_{c\bar b}$ tetraquark states are represented $\varepsilon^\delta$ and, $\varepsilon^{\theta}$ and $\varepsilon^\tau$ is the polarization vector of the photon.  Here, $G_1(Q^2)$, $G_2(Q^2)$ and $G_3(Q^2)$ are invariant form factors,  with  $Q^2=-q^2$.

Employing Eqs.~(\ref{edmn04})-(\ref{edmn06}), the correlator takes the form,
%\begin{widetext}
%
\begin{align}
\label{edmn09}
 \Pi_{\mu\nu}^{Had}(p,q) &=  \frac{\varepsilon_\rho \, \lambda_{Z_{c\bar b}}^2}{ [m_{Z_{c\bar b}}^2 - (p+q)^2][m_{Z_{c\bar b}}^2 - p^2]}
 %\nonumber\\
 %&
 \bigg\{G_1(Q^2)(2p+q)_\rho\bigg[g_{\mu\nu}-\frac{p_\mu p_\nu}{m_{Z_{c\bar b}}^2}
 %\nonumber\\
 %&
 -\frac{(p+q)_\mu (p+q)_\nu}{m_{Z_{c\bar b}}^2}+\frac{(p+q)_\mu p_\nu}{2m_{Z_{c\bar b}}^4}\nonumber\\
 & \times (Q^2+2m_{Z_{c\bar b}}^2)
 \bigg]
 + G_2 (Q^2) \bigg[q_\mu g_{\rho\nu}  
 %\nonumber\\
 %&
 - q_\nu g_{\rho\mu}-
\frac{p_\nu}{m_{Z_{c\bar b}}^2}  \big(q_\mu p_\rho - \frac{1}{2}
Q^2 g_{\mu\rho}\big) 
%\nonumber\\
%&
+
\frac{(p+q)_\mu}{m_{Z_{c\bar b}}^2}  \big(q_\nu (p+q)_\rho+ \frac{1}{2}
Q^2 g_{\nu\rho}\big) 
\nonumber\\
&-  
\frac{(p+q)_\mu p_\nu p_\rho}{m_{Z_{c\bar b}}^4} \, Q^2
\bigg]
%\nonumber\\
%&
-\frac{G_3(Q^2)}{m_{Z_{c\bar b}}^2}(2p+q)_\rho \bigg[
q_\mu q_\nu -\frac{p_\mu q_\nu}{2 m_{Z_{c\bar b}}^2} Q^2 
%\nonumber\\
%&
+\frac{(p+q)_\mu q_\nu}{2 m_{Z_{c\bar b}}^2} Q^2
-\frac{(p+q)_\mu q_\nu}{4 m_{Z_{c\bar b}}^4} Q^4\bigg]
\bigg\}\,.
\end{align}

%\end{widetext}

The  magnetic dipole ($F_M(Q^2)$) and quadrupole ($F_{\cal D}(Q^2)$) form factors  are described  in terms of $G_1(Q^2)$, $G_2(Q^2)$  and $G_3(Q^2)$ form factors as:  
\begin{align}
\label{edmn07}
&F_M(Q^2) = G_2(Q^2)\,,\nonumber \\
&F_{\cal D}(Q^2) = G_1(Q^2)-G_2(Q^2)+(1+\tau) G_3(Q^2)\,,
\end{align}
where $\tau=Q^2/4 m_{Z_{c\bar b}}^2$ with $Q^2=-q^2$. In the static limit, $Q^2 = 0 $, the $F_M(Q^2=0)$ and $F_{\cal D}(Q^2=0)$ form factors are related to the magnetic moment ($\mu$), and quadrupole moment (${\cal D}$) as follows
\begin{align}
\label{edmn08}
&e F_M(0) = 2 m_{Z_{c\bar b}} \mu \,, \nonumber\\
&e F_{\cal D}(0) = m_{Z_{c\bar b}}^2 {\cal D}\,.
\end{align}

In  QCD representation, the correlator in Eq.~(\ref{edmn01}), is evaluated concerning the QCD degrees of freedom in deep Euclidean region where $p^2 <<0$ and $(p+q)^2 <<0$. To do this,  we insert the interpolating currents in the correlator and contract the heavy and light quark fields utilizing Wick's theorem. As a result of these steps,  %the correlator takes the following form
we get following expressions for the $Z_{c\bar b}$ states
%\begin{widetext}

\begin{align}
\Pi _{\mu \nu }^{\mathrm{QCD}-Mol}(p,q)&=i\int d^{4}xe^{ip\cdot x} \langle 0 \mid \  \mathrm{Tr}%
\big[ \gamma _{\mu }{S}_{q}^{bb^{\prime }}(x)\gamma _{\nu
}  
%\nonumber\\
%& \times 
S_{b}^{b^{\prime }b}(-x)\big]
\mathrm{Tr}\big[ \gamma _{5}{S}_{c}^{aa^{\prime
}}(x)\gamma _{5}S_{q}^{a^{\prime  }a}(-x)\big] 
%\nonumber\\
%& \times
 \mid 0 \rangle_{\gamma} ,  \label{eq:QCDSide}
\end{align}%

\begin{align}
\Pi _{\mu \nu }^{\mathrm{QCD}-Di}(p,q)&=i\int d^{4}xe^{ip\cdot x} \langle 0 \mid \bigg\{  \mathrm{Tr}%
\Big[ \gamma _{\mu }\widetilde{S}_{b}^{b^{\prime }b}(-x)\gamma _{\nu
}   
%\nonumber\\
%&\times 
S_{q}^{a^{\prime }a}(-x)\Big]    \mathrm{Tr}\Big[ \gamma _{5} \widetilde{S}_{q}^{aa^{\prime
}}(x)\gamma _{5}S_{c}^{bb^{\prime }}(x)\Big] \nonumber\\
&+\mathrm{Tr}\Big[ \gamma
_{\mu }\widetilde{S}_{b}^{a^{\prime }b}(-x) \gamma _{\nu }S_{q}^{b^{\prime }a}(-x)\Big]
%\nonumber\\
%& \times 
\mathrm{Tr}%
\Big[ \gamma _{5}\widetilde{S}_{q}^{aa^{\prime }}(x)\gamma
_{5}S_{c}^{bb^{\prime }}(x)\Big]  \nonumber \\
&+\mathrm{Tr}\Big[ \gamma _{\mu }\widetilde{S}_{b}^{b^{\prime
}a}(-x)\gamma _{\nu }S_{q}^{a^{\prime }b}(-x)\Big] 
%\nonumber\\
%& \times 
\mathrm{Tr}\Big[
\gamma _{5}\widetilde{S}_{q}^{aa^{\prime }}(x)\gamma _{5}S_{c}^{bb^{\prime
}}(x)\Big]  \notag \\
& +\mathrm{Tr}\Big[ \gamma _{\mu }\widetilde{S}_{b}^{a^{\prime
}a}(-x)\gamma _{\nu }S_{q}^{b^{\prime }b}(-x)\Big]
%\nonumber\\
%& \times 
\mathrm{Tr}\Big[
\gamma _{5}\widetilde{S}_{q}^{aa^{\prime }}(x) \gamma _{5}S_{c}^{bb^{\prime
}}(x)\Big] \bigg\} \mid 0 \rangle_{\gamma} ,  \label{eq:QCDSide2}
\end{align}%
%\end{widetext}
 where
\begin{equation*}
\widetilde{S}_{Q(q)}^{ij}(x)=CS_{Q(q)}^{ij\mathrm{T}}(x)C,
\end{equation*}%
with $C$ and $T$ being the charge conjugation and transpose of the
operator, respectively. Here, $S_{q}(x)$ and $S_{Q}(x)$ represent the full light and heavy quark propagators.  Throughout our calculations, we use the x-space expressions for the light and heavy quark 
propagators~\cite{Yang:1993bp, Belyaev:1985wza}:
\begin{align}
\label{edmn12}
S_{q}(x)&=i \frac{{\xslash}}{2\pi ^{2}x^{4}} 
- \frac{\langle \bar qq \rangle }{12} \Big(1-i\frac{m_{q} \xslash}{4}   \Big)
- \frac{ \langle \bar qq \rangle }{192}m_0^2 x^2 
%\nonumber\\
%& \times 
\Big(1 -i\frac{m_{q} \xslash}{6}   \Big)
-\frac {i g_s }{32 \pi^2 x^2} ~G^{\mu \nu} (x) \Big[\rlap/{x} 
\sigma_{\mu \nu} +  \sigma_{\mu \nu} \rlap/{x}
 \Big],
\end{align}
%and
%%
\begin{align}
\label{edmn13}
S_{Q}(x)&=\frac{m_{Q}^{2}}{4 \pi^{2}} \Bigg[ \frac{K_{1}\Big(m_{Q}\sqrt{-x^{2}}\Big) }{\sqrt{-x^{2}}}
+i\frac{{\xslash}~K_{2}\Big( m_{Q}\sqrt{-x^{2}}\Big)}
{(\sqrt{-x^{2}})^{2}}\Bigg]
%\nonumber\\
%&
-\frac{g_{s}m_{Q}}{16\pi ^{2}} \int_0^1 dv\, G^{\mu \nu }(vx)\Bigg[ \big(\sigma _{\mu \nu }{\xslash}
  +{\xslash}\sigma _{\mu \nu }\big)\nonumber\\
  & \times \frac{K_{1}\Big( m_{Q}\sqrt{-x^{2}}\Big) }{\sqrt{-x^{2}}} 
+2\sigma_{\mu \nu }K_{0}\Big( m_{Q}\sqrt{-x^{2}}\Big)\Bigg],
\end{align}%
where $\langle \bar qq \rangle$ is light-quark  condensate, $m_0$ is defined through  $\langle 0 \mid \bar  q\, g_s\, \sigma_{\alpha\beta}\, G^{\alpha\beta}\, q \mid 0 \rangle = m_0^2 \,\langle \bar qq \rangle $, $G^{\mu\nu}$ is the gluon field strength tensor, $v$ is line variable %,  $\sigma_{\mu\nu}= \frac{i}{2}[\gamma_\mu, \gamma_\nu]$ 
and $K_i$'s are modified Bessel functions of the second kind.  
The first term of the light and heavy quark propagators correspond to the perturbative or free part and the rest belong to the interacting parts. 
The correlator in QCD representation includes two different contributions: perturbative and nonperturbative.  How perturbative and nonperturbative contributions are calculated in the analysis is given in Appendix A using Eq. (\ref{eq:QCDSide}) as an example.
The QCD degrees of freedom representation of the correlator can be obtained in terms of the quark-gluon properties via the photo DAs and after performing an integration over x, the expression of the correlator in the momentum representation can be calculated straightforwardly. 

To determine LCSR for the magnetic and quadrupole moments, we perform double Borel transformation and continuum subtraction to suppress the higher states and continuum effects. As a result, we get
\begin{align}\label{MMmol}
 &\mu_{Z_{c\bar b}}^{Mol}\,\, \lambda_{Z_{c\bar b}}^{2-Mol}  = e^{\frac{m_{Z_{c\bar b}}^2}{M^2}} \,\, \Delta_1^{QCD}(M^2,s_0),\\
 %\end{align}
 %\begin{align}\label{MMDi}
 &\mu_{Z_{c\bar b}}^{Di}\,\, \lambda_{Z_{c\bar b}}^{2-Di}  = e^{\frac{m_{Z_{c\bar b}}^2}{M^2}} \,\, \Delta_2^{QCD}(M^2,s_0),
\end{align}
\begin{align}\label{QMmol}
 &\mathcal{D}_{Z_{c\bar b}}^{Mol} \,\, \lambda_{Z_{c\bar b}}^{2-Mol} = m^2_{Z_{c \bar b}}e^{\frac{m_{Z_{c\bar b}}^2}{M^2}} \,\,\Delta_3^{QCD}(M^2,s_0),\\
%\end{align}
%\begin{align}\label{QMDi}
& \mathcal{D}_{Z_{c\bar b}}^{Di} \,\, \lambda_{Z_{c\bar b}}^{2-Di} = m^2_{Z_{c \bar b}}e^{\frac{m_{Z_{c\bar b}}^2}{M^2}} \,\,\Delta_4^{QCD}(M^2,s_0). \label{QMDi}
\end{align}

For the sake of simplicity, only the explicit expressions of the $\Delta_1^{QCD}(M^2,s_0)$ and $\Delta_2^{QCD}(M^2,s_0)$  functions are presented in Appendix B.

%\end{widetext}

\section{Numerical illustrations and Discussions}\label{numerical}

 In this section, we perform the LCSR analyses for the magnetic and quadrupole moments of $Z_{c\bar b}$ tetraquark states. We use the values of masses, residues and various QCD condensates as follows: $m_u=m_d=0$,  $m_c = 1.67 \pm 0.07\,\mbox{GeV}$, $m_b = 4.78 \pm 0.06\,\mbox{GeV}$, $m_{Z_{c\bar b}}^{Mol} = 6.85 \pm 0.15\,\mbox{GeV}$, $m_{Z_{c\bar b}}^{Di} = 7.06 \pm 0.78\,\mbox{GeV}$, $\lambda_{Z_{c\bar b}}^{Mol} = 0.036 \pm 0.011\,\mbox{GeV}^5$ \cite{Albuquerque:2012rq}, $\lambda_{Z_{c\bar b}}^{Di} = 0.024 \pm 0.08\,\mbox{GeV}^5$ \cite{Agaev:2017uky}   $m_0^{2} = 0.8 \pm 0.1 \,\mbox{GeV}^2$ \cite{Ioffe:2005ym},   $\langle \bar uu\rangle = 
\langle \bar dd\rangle=(-0.24 \pm 0.01)^3\,\mbox{GeV}^3$ \cite{Ioffe:2005ym},  
$\langle g_s^2G^2\rangle = 0.88~ \mbox{GeV}^4$~\cite{Nielsen:2009uh},  $f_{3\gamma}=-0.0039~\mbox{GeV}^2$~\cite{Ball:2002ps} and $\chi=-2.85 \pm 0.5~\mbox{GeV}^{-2}$~\cite{Rohrwild:2007yt}.   
To proceed with the numerical calculations distribution amplitudes of the photon are needed. The explicit expressions of the photon distribution amplitudes and numerical values of input parameters are borrowed from Ref.~\cite{Ball:2002ps}.

There are two free parameters in Eqs. (\ref{MMmol})-(\ref{QMDi}), the Borel mass parameter ($M^2$) and the continuum threshold ($s_0$). By taking into consideration that the magnetic and quadrupole moments are physical observables they should be less dependent on the free parameters $M^2$, and $s_0$.    
 To obtain reliable working regions for these free parameters, we use three criteria to constrain them: The convergence of the operator product expansion (OPE),  the pole contribution ($PC$), and the magnetic and quadrupole moments dependence on $M^2$ and $s_0$.  
In the LCSR analysis, the $PC$ requires that it should exceed $20 \%$ of the total contributions, which is typical for the multiquark states. We also demand that the series of light-cone expansion converge and contributions of the higher twist and higher condensate terms are less than $10 \%$ of the total contribution. Considering these constraints working intervals of these helping parameters have been determined as follows:
\begin{align}
 58.0~\mbox{GeV}^2 \leq s_0 \leq 60.0~\mbox{GeV}^2, 
 \nonumber\\
 %\nonumber\\
 7.5~\mbox{GeV}^2 \leq M^2 \leq 9.5~\mbox{GeV}^2.
\end{align}

\begin{widetext}
 
\begin{figure}[t]
\centering
  \includegraphics[width=0.4\textwidth]{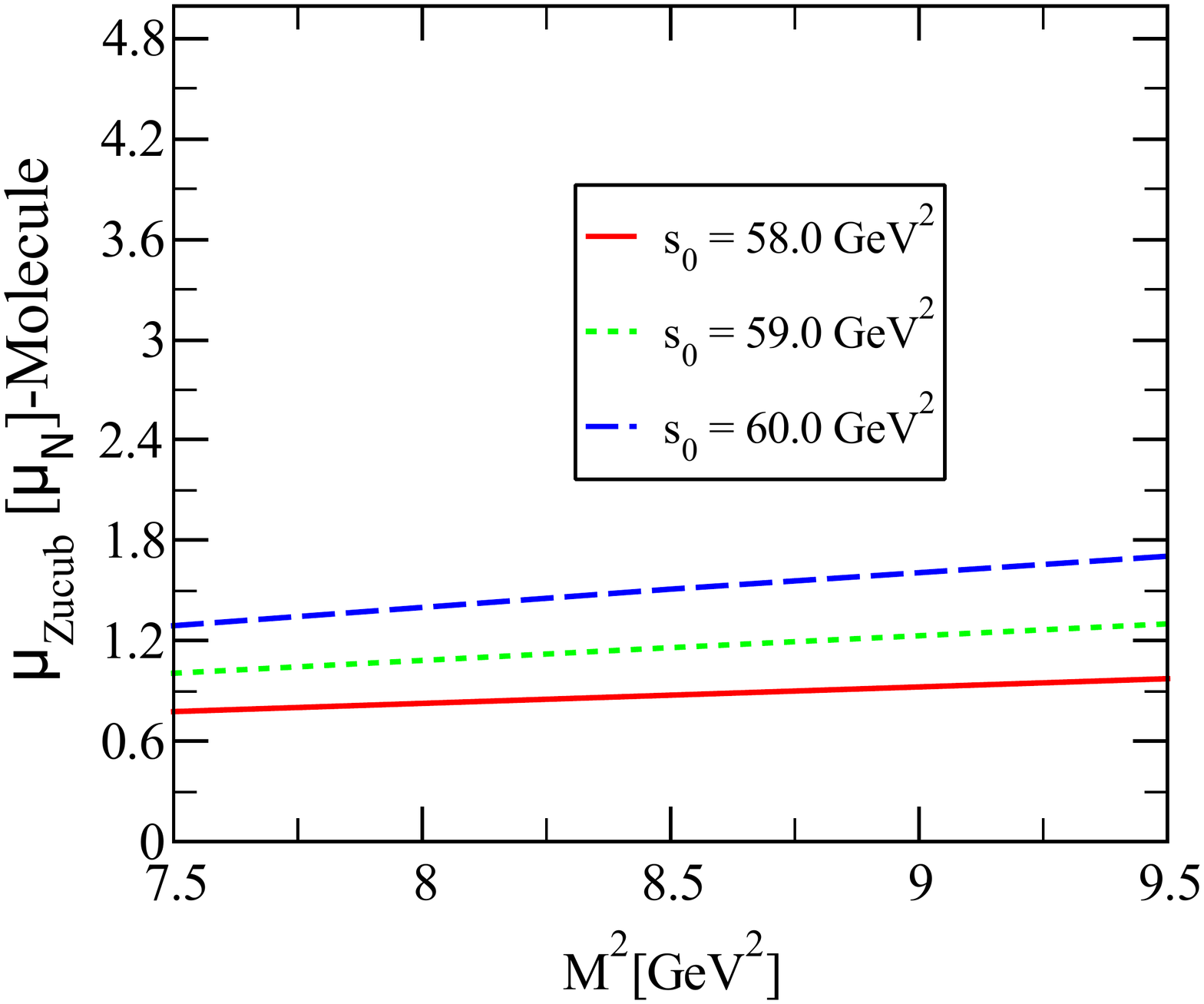} 
  \includegraphics[width=0.4\textwidth]{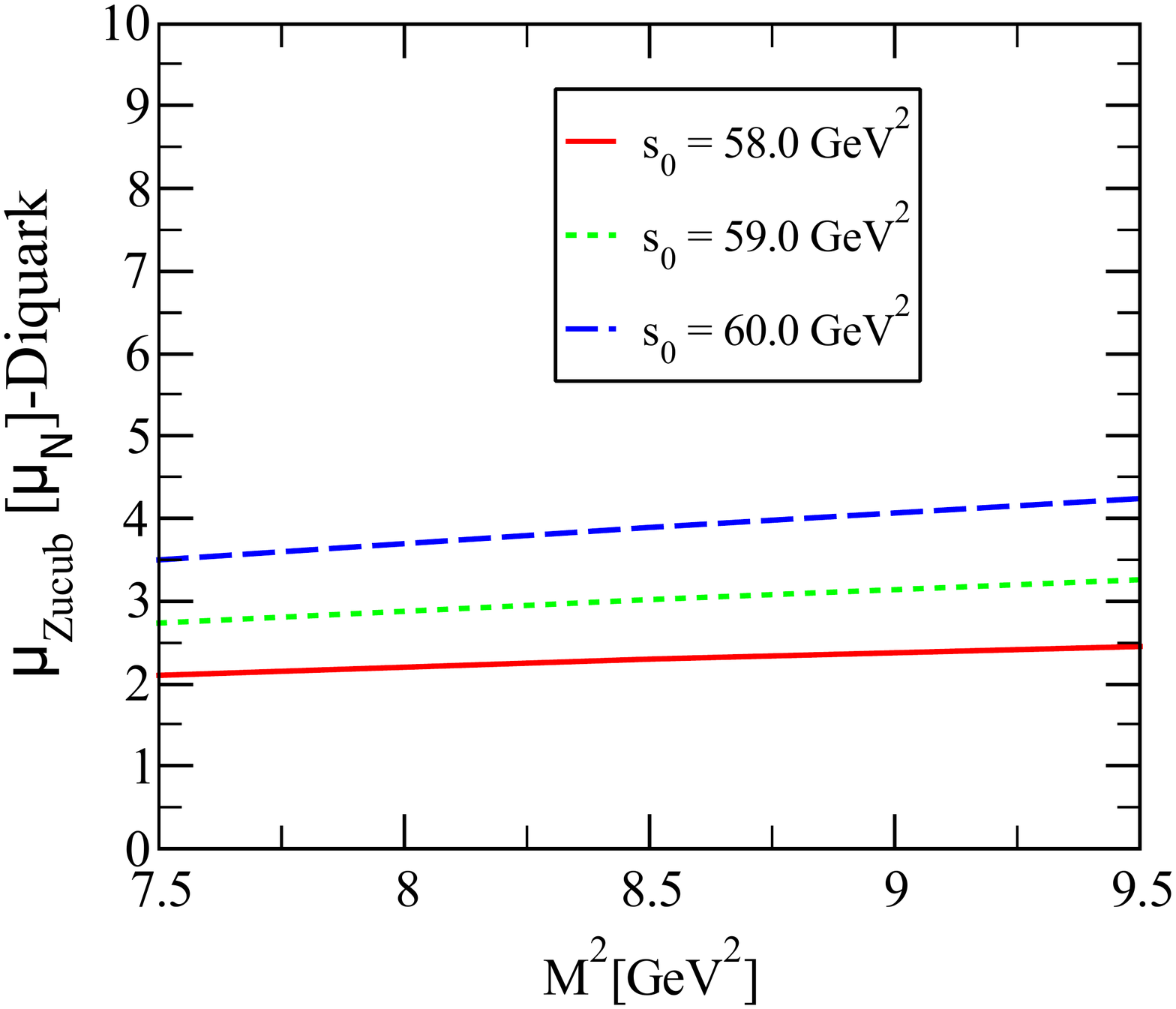} \\
 \includegraphics[width=0.4\textwidth]{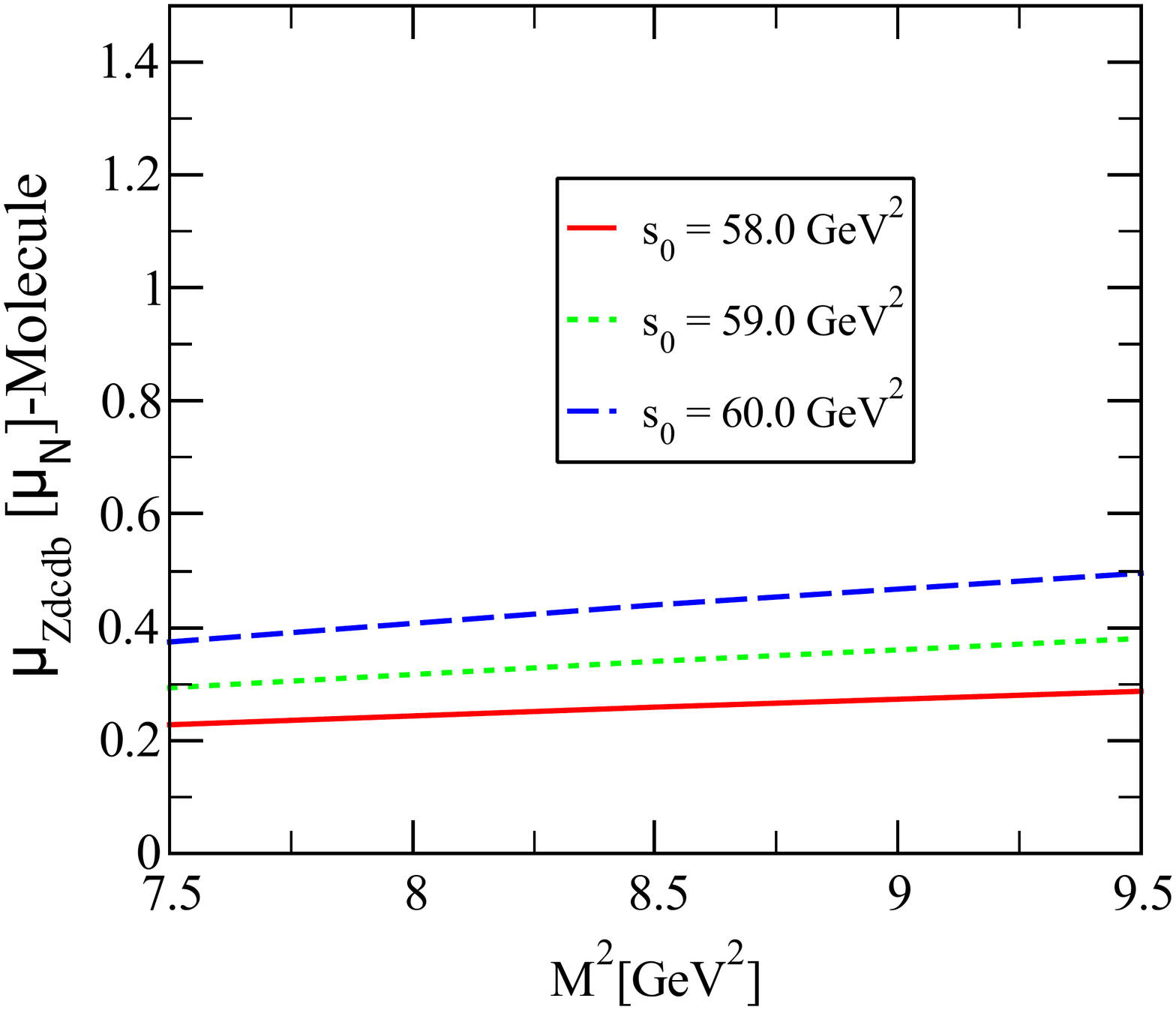}
 \includegraphics[width=0.4\textwidth]{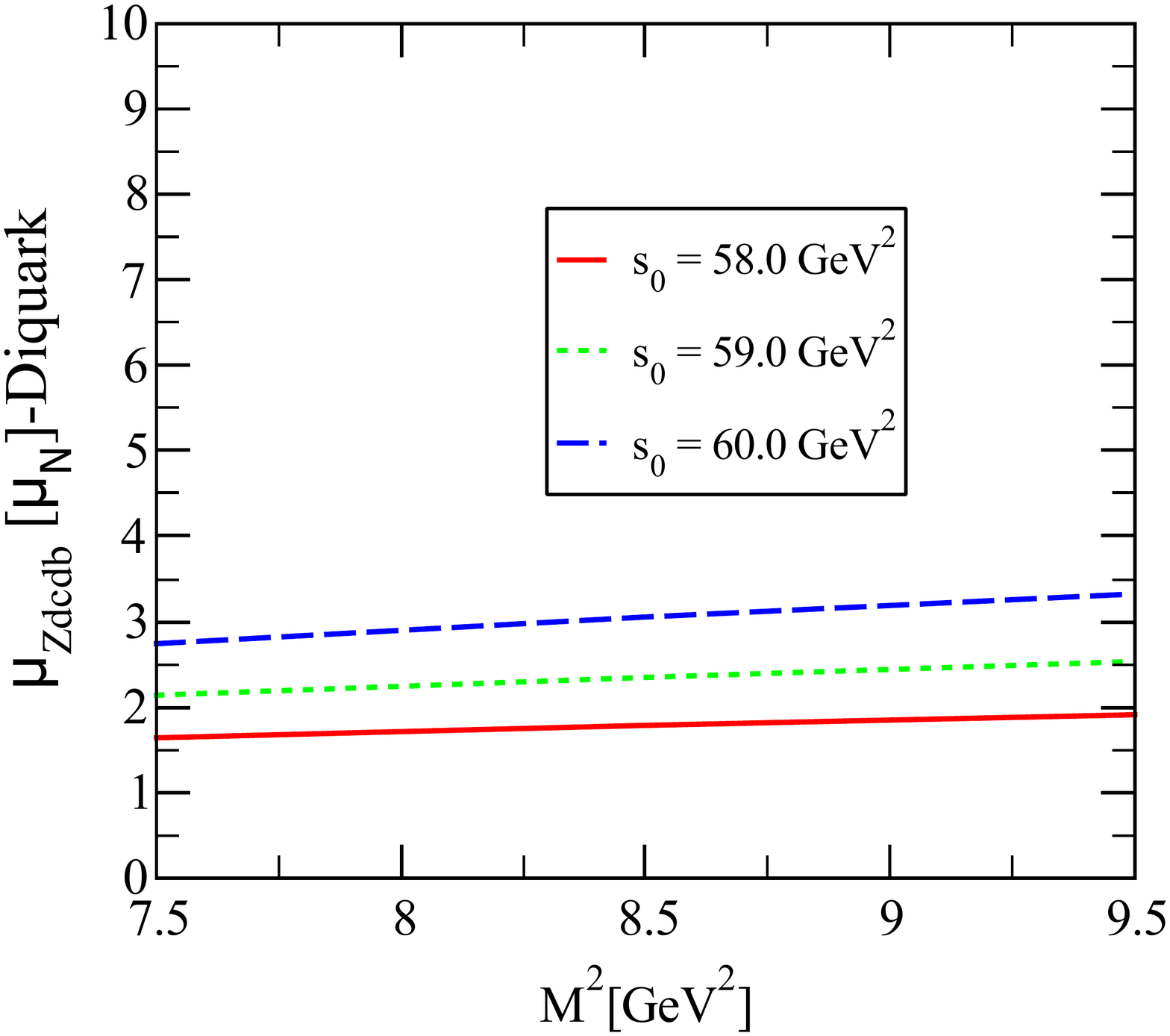}
 \caption{ The dependence of magnetic moment of the $Z_{c\bar b}$ tetraquark states on $M^{2}$  at three fixed values of $s_0$.}
 \label{figMsq}
  \end{figure}

  \end{widetext}

Our numerical computations indicate that by taking into account these working intervals for the helping parameters, the magnetic and quadrupole moments of the $Z_{c\bar b}$ tetraquark states $  PC$ varies within the interval $30\%\leq PC \leq 59\%$ corresponding to the upper and lower limits of the Borel mass parameter. 
When we analyze the OPE convergence, we see that the contribution of the higher twist and higher dimensional terms in OPE is  $3 \%$ of the total and the series shows good convergence.
As an example, in Fig. \ref{figMsq}, we also show the variation of the extracted magnetic moment of $Z_{c\bar b}$ tetraquark states with the Borel mass $M^2$ using three fixed values of the continuum threshold $s_0$. It can be seen from this figure that the magnetic moment indicates a weak dependence on $M^2$ in its working region. Though the magnetic moment of $Z_{c\bar b}$ tetraquark states show some dependence on $s_0$, it remains inside the limits allowed by the method and constitutes the main parts of the uncertainties.

We have determined all necessary parameters to specify the numerical values for the magnetic and quadrupole moments of  $Z_{c\bar b}$ tetraquark states.
 As a result of our comprehensive numerical calculations, the magnetic moments of $Z_{c\bar b}$ tetraquark states are given in Table \ref{table}. %finally found to have the values
 It should be noted here that in numerical calculations we take into account the uncertainties in the input parameters, uncertainties entering into the photon DAs, as well as uncertainties because of the variations of Borel mass parameter $M^2$ and continuum threshold $s_0$. We would like to point out that roughly 18$\%$ of the errors in the numerical results are due to the mass of tetraquarks, 15$\%$ belongs to the residue of tetraquarks, 30$\%$ belongs to $s_0$, 7$\%$ belongs to $M^2$, 13$\%$ belongs to photon DAs and the remaining 17$\%$ corresponds to other input parameters.
 
 \begin{table}[htp]
	\addtolength{\tabcolsep}{11pt}
	\caption{Magnetic  moments  of the $Z_{c\bar b}$ states (in units of nuclear
magnetons $\mu_N$).}
	\label{table}
	\begin{center}
\begin{tabular}{lccccc}
	   \hline\hline
	   \\
	   Picture&  $\mu_{Z_{{uc \bar u \bar b}}}$	& $\mu_{Z_{dc \bar d \bar b}}$		   \\
	  % \\
	  \\
	   \hline\hline
	  \\
	   Diquark&   $ 3.05^{+1.19}_{-0.95}$     &$ 2.38^{+0.95}_{-0.75}$ 	\\
	 %  \\
	 \\
	   \hline\hline
	   \\
	   Molecule&  $1.18^{+0.52}_{-0.40}$     &$0.32^{+0.18}_{-0.10}$ 	
	   \\
	   \\
	   \hline\hline
\end{tabular}
\end{center}
\end{table}

%\begin{align}
%  \mu_{Z_{{uc \bar u \bar b}}}^{Mol}&=1.18^{+0.52}_{-0.40}~\mu_N, ~~~~\mu_{Z_{{uc \bar u \bar b}}}^{Di}=3.05^{+1.19}_{-0.95}~\mu_N,\\
%  \nonumber\\
%  \mu_{Z_{dc \bar d \bar b}}^{Mol}&=0.32^{+0.18}_{-0.10}~\mu_N,   ~~~~ \mu_{Z_{dc \bar d \bar b}}^{Di}=2.38^{+0.95}_{-0.75}~\mu_N,
% \end{align}%\\
%where $\mu_N$ is nuclear magneton.

The magnitude of the results obtained for the magnetic moments also gives the possibility to be measured experimentally.  %As can be seen from the magnetic moment results, these results are large enough to be measured experimentally. 
It follows from these results that the magnetic moments of $B_c$-like ground state tetraquarks are large enough to be measured in future experiments.
When we compare the numerical results of the two pictures above, we observe that states with the same quantum numbers have different magnetic moments, which clearly shows the magnetic moment strongly depends on the structure of the hadron. This implies that any possible experimental measurement of magnetic moments can help us understand the internal structure of these states.
Different magnetic moments will affect both the total and differential cross sections in the photo- or electro-production of $B_c$-like tetraquarks. Thus, information on the $B_c$-like tetraquarks magnetic moments will help us unveil the mysterious curtain over the $B_c$-like tetraquarks and deepen our understanding of the underlying quark structure and dynamics. Our results on magnetic moments of the $B_c$-like ground state tetraquarks may be checked in the framework of the alternative phenomenological models.
Hopefully, the examinations given in this study will be helpful to an experimental search of them, which will be an interesting research subject.

To our best knowledge, this is the first study in the literature committed to the investigation of $B_c$-like tetraquark magnetic and quadrupole moments. Hence, theoretical predictions or experimental data  are not yet existing to compare them with our numerical values. 
However, we give the magnetic moment results of the standard $B_c^*$ meson so that the reader can better understand the difference in the results obtained for the  $B_c$-like ground state tetraquarks. The magnetic moments of $B_c^*$  meson have been extracted using the Bag model ~\cite{Bose:1980vy}, extended Bag model \cite{Simonis:2018rld} and  Blankenbecler-Sugar (BSLT) equation \cite{Lahde:2002wj}.  The obtained results are given as $\mu_{B_c^*} = 0.20\, \mu_N $,  $\mu_{B_c^*} = 0.38\, \mu_N $ and  $\mu_{B_c^*} = 0.426\, \mu_N $ for the Bag model, extended Bag model, and BSLT, respectively. We would like to point out that only magnetic moment results for vector $B_c^*$ mesons have been calculated in the literature and therefore these results are presented. As we mentioned above, it is given only to make the results understandable for the reader. 

We also present numerical results of quadrupole moments of $B_c$-like tetraquark states, and their numerical values are given in Table \ref{table2}.  
We can notice that just like the magnetic moment results, the quadrupole moment results obtained using two different configurations are different from each other.  The quadrupole moments of the $B_c$-like tetraquark states indicate non-spherical charge distributions. 
The sign of quadrupole moments are negative for  $Z_{{uc \bar u \bar b}}$ tetraquark states and positive for $Z_{dc \bar d \bar b}$ tetraquark states, which correspond to the oblate and prolate charge distributions, respectively.

%\begin{align}
%  \mathcal{D}_{Z_{{uc \bar u \bar b}}}^{Di}&=-(1.40^{+0.40}_{-0.40})\times 10^{-2}~fm^2, \nonumber\\\mathcal{D}_{Z_{{uc \bar u \bar b}}}^{Mol} &=-(0.40^{+0.10}_{-0.10})\times 10^{-2}~fm^2,%\\
  %\nonumber\\
%  \end{align}
%  \begin{align}
%  \mathcal{D}_{Z_{dc \bar d \bar b}}^{Di}&=(0.70^{+0.30}_{-0.30})\times 10^{-2}~fm^2,\nonumber\%\
%  \mathcal{D}_{Z_{dc \bar d \bar b}}^{Mol} &=(0.20^{+0.05}_{-0.05})\times 10^{-2}~fm^2.
% \end{align}%\\
%where $\mu_N$ is nuclear magneton.
%
\begin{table}[htp]
	\addtolength{\tabcolsep}{11pt}
	\caption{Quadrupole  moments  of the $Z_{c\bar b}$ states (in units of $fm^2$).}
	\label{table2}
	\begin{center}
\begin{tabular}{lccccc}
	   \hline\hline
	   \\
	   Picture&  $\mathcal{D}_{Z_{{uc \bar u \bar b}}}(\times 10^{-2})$	& $\mathcal{D}_{Z_{dc \bar d \bar b}}(\times 10^{-2})$		   \\
	   \\
	   \hline\hline
	  \\
	   Diquark&   $ -1.40^{+0.40}_{-0.40}$     &$ 0.70^{+0.30}_{-0.30}$ 	\\
	   \\
	   \hline\hline
	   \\
	   Molecule&  $-0.40^{+0.10}_{-0.10}$     &$0.20^{+0.05}_{-0.05}$ 	
	   \\
	  \\
	   \hline\hline
\end{tabular}
\end{center}
\end{table}

\section{summary and Outlook}\label{sum}

We have employed the molecular and diquark-antidiquark tetraquark pictures to extract magnetic and quadrupole moments of the $B_c$-like ground state tetraquarks with the QCD light-cone sum rules with quantum numbers $J^P = 1^+$.  
Comparing the numerical results of the above two configurations, we notice that the magnetic moments of the $B_c$-like tetraquark states with the same quantum numbers differ significantly from each other, which can be used to identify the underlying structure of these states. We have also extracted the quadrupole moments of these states.  The quadrupole moments of the $B_c$-like tetraquark states indicate non-spherical charge distributions.

The magnetic moment of hadrons is an essential ingredient in calculations of the photo- and electro-production cross sections and may be obtained from the experiments in the future. 
With the increased luminosity in future runs, spectroscopic parameters and magnetic moments of $B_c$-like tetraquark states may be extracted from experimental facilities, which may help distinguish different theoretical approaches and deepen our understanding of the underlying dynamics governing their formations.
It will also be important to determine the branching ratios of the different decay modes and decay channels of the $B_c$-like tetraquark states. Furthermore, the study of the $B_c$-like tetraquark states in other theoretical models can also be very interesting.

%\newpage

\begin{widetext}
   \section*{Appendix A: perturbative and nonperturbative contributions of QCD side of the correlation function}
   
   In this appendix, we will show in a short example how the perturbative and nonperturbative contributions, which appear in the analysis of the QCD side of the correlation function, are calculated.
   \begin{align}
\Pi _{\mu \nu }^{\mathrm{QCD}-Mol}(p,q)&=i\int d^{4}xe^{ip\cdot x} \langle 0 \mid \  \mathrm{Tr}%
\big[ \gamma _{\mu }{S}_{q}^{bb^{\prime }}(x)\gamma _{\nu
}  
%\nonumber\\
%& \times 
S_{b}^{b^{\prime }b}(-x)\big]
\mathrm{Tr}\big[ \gamma _{5}{S}_{c}^{aa^{\prime
}}(x)\gamma _{5}S_{q}^{a^{\prime  }a}(-x)\big] 
%\nonumber\\
%& \times
 \mid 0 \rangle_{\gamma},  \label{new:QCDSide}
\end{align}%
   
   As we mentioned above, the correlator in Eq. (\ref{new:QCDSide}) includes  different types of contributions: the photon can be emitted both perturbatively or
non-perturbatively. 

Practically, perturbative contributions, the photon interacts with one of the quarks, can be computed by replacing one of the light or heavy-quark propagators in Eq.~(\ref{new:QCDSide}) by
\begin{align}
\label{sfree}
S^{free} \rightarrow \int d^4y\, S^{free} (x-z)\,\rlap/{\!A}(z)\, S^{free} (z)\,,
\end{align}
the remaining three propagators are taken as full quark propagators. The QCD light-cone sum rule calculations are usually done on a fixed-point gauge is also known as Fock-Schwinger gauge. The most important reason for using this gauge is to express the external background field according to the field strength tensor and also the use of this gauge is preferred because it provides some convenience in calculations. For the electromagnetic field, it is characterized by $x_\mu A^\mu =0$.  In this gauge, the external electromagnetic potential is given by 
\begin{align}
\label{AAA}
 &A_\alpha (z) = -\frac{1}{2} F_{\alpha\beta}z^\beta 
   = -\frac{1}{2} (\varepsilon_\alpha q_\beta-\varepsilon_\beta q_\alpha)\,z^\beta.
\end{align}
Equation (\ref{AAA}) is inserted into Eq. (\ref{sfree}), we obtain
 \begin{align}
  S^{free} \rightarrow -\frac{1}{2} (\varepsilon_\alpha q_\beta-\varepsilon_\beta q_\alpha)
  \int\, d^4z \,z^{\beta}\, 
  S^{free} (x-z)\,\gamma_{\alpha}\,S^{free} (z)\,,
 \end{align}

After some computations for $S_q^{free}$ and $S_Q^{free}$, their final form becomes:
\begin{eqnarray}\label{sfreepert}
&& S_q^{free}(x)=\frac{e_q}{32 \pi^2 x^2}\Big(\varepsilon_\alpha q_\beta-\varepsilon_\beta q_\alpha\Big)
 \Big(\xslash\sigma_{\alpha \beta}+\sigma_{\alpha\beta}\xslash\Big),\nonumber\\
&& S_Q^{free}(x)=-i\frac{e_Q m_Q}{32 \pi^2}
\Big(\varepsilon_\alpha q_\beta-\varepsilon_\beta q_\alpha\Big)
\Big[2\sigma_{\alpha\beta}K_{0}\Big( m_{Q}\sqrt{-x^{2}}\Big)
 +\frac{K_{1}\Big( m_{Q}\sqrt{-x^{2}}\Big) }{\sqrt{-x^{2}}}
 \Big(\xslash\sigma_{\alpha \beta}+\sigma_{\alpha\beta}\xslash\Big)\Big].
\end{eqnarray}

Equation (\ref{sfreepert}) is inserted into Eq. (\ref{new:QCDSide}), and as a result these manipulations for the perturbative contributions we get 

   \begin{align}
\Pi _{\mu \nu -Pert }^{\mathrm{QCD}-Mol}(p,q)&=i\int d^{4}xe^{ip\cdot x} \, \mathrm{Tr}%
\big[ \gamma _{\mu }{S}_{q}^{free}(x)\gamma _{\nu
}  
S_{b}^{b^{\prime }b}(-x)\big]
\mathrm{Tr}\big[ \gamma _{5}{S}_{c}^{aa^{\prime
}}(x)\gamma _{5}S_{q}^{a^{\prime  }a}(-x)\big] \delta^{bb^{\prime }}
\nonumber\\
& 
+
\mathrm{Tr}%
\big[ \gamma _{\mu }{S}_{q}^{b^{\prime }b}(x)\gamma _{\nu
}  
S_{b}^{free}(-x)\big]
\mathrm{Tr}\big[ \gamma _{5}{S}_{c}^{aa^{\prime
}}(x)\gamma _{5}S_{q}^{a^{\prime  }a}(-x)\big]\delta^{bb^{\prime }}
\nonumber\\
&
+
\mathrm{Tr}%
\big[ \gamma _{\mu }{S}_{q}^{b^{\prime }b}(x)\gamma _{\nu
}   
S_{b}^{b^{\prime }b}(-x)\big]
\mathrm{Tr}\big[ \gamma _{5}{S}_{c}^{free}(x)\gamma _{5}S_{q}^{a^{\prime  }a}(-x)\big]\delta^{aa^{\prime }}\nonumber\\
&
+
\mathrm{Tr}%
\big[ \gamma _{\mu }{S}_{q}^{b^{\prime }b}(x)\gamma _{\nu
}  
%\nonumber\\
%& \times 
S_{b}^{b^{\prime }b}(-x)\big]
\mathrm{Tr}\big[ \gamma _{5}{S}_{c}^{aa^{\prime
}}(x)\gamma _{5}S_{q}^{free}(-x)\big]\delta^{a^{\prime }a}.  \label{new:QCDSide1}
\end{align}%

It should be noted that all possibilities have been considered in the above equation. In the first line of Eq. (\ref{new:QCDSide1}), the photon interacts perturbatively with one of the light quark propagators, while the remaining three quark propagators are taken as full. Similarly, in the second line of Eq. (\ref{new:QCDSide1}), the photon interacts perturbatively with one of the heavy quark propagators, while the other propagators are taken as full, and so on.

Nonperturbative contributions, the photon is radiated at long distances, can be computed by replacing one of the light quark propagators in Eq.~(\ref{new:QCDSide}) by 
\begin{align}
\label{neweq}
S_{\alpha\beta}^{ab} \rightarrow -\frac{1}{4} (\bar{q}^a \Gamma_i q^b)(\Gamma_i)_{\alpha\beta},
\end{align}
where $\Gamma_i = I, \gamma_5, \gamma_\mu, i\gamma_5 \gamma_\mu, \sigma_{\mu\nu}/2$ and the remaining three propagators are considered as full quark propagators including the perturbative as well as the nonperturbative contributions. In the second case, the correlator takes the form,
\begin{eqnarray}\label{QCDES}
\Pi _{\mu \nu-Nonpert }^{\mathrm{QCD}-Mol}(p,q)&=&-\frac{i}{4}\int d^{4}xe^{ip\cdot x} \langle 0 \mid \ \Bigg\{ \mathrm{Tr}%
\big[ \gamma _{\mu }\Gamma_i\gamma _{\nu
}  
S_{b}^{b^{\prime }b}(-x)\big] \Big(\bar q ^b (x) \Gamma_i  q^{b^{\prime }}(0)\Big) \,
\mathrm{Tr}\big[ \gamma _{5}{S}_{c}^{aa^{\prime
}}(x)\gamma _{5}S_{q}^{a^{\prime  }a}(-x)\big]  
\nonumber\\
&&+
\mathrm{Tr}%
\big[ \gamma _{\mu }{S}_{q}^{bb^{\prime }}(x)\gamma _{\nu
}  
%\nonumber\\
%& \times 
S_{b}^{b^{\prime }b}(-x)\big]
\mathrm{Tr}\big[ \gamma _{5}{S}_{c}^{aa^{\prime
}}(x)\gamma _{5}\Gamma_i \big] \Big(\bar q ^a (x) \Gamma_i  q^{a^{\prime }}(0)\Big) \Bigg\}
 \mid 0 \rangle_{\gamma}, 
\end{eqnarray}%

 By replacing one of the light quark propagators and using the expression $  \bar q^a(x)\Gamma_i q^{a'}(0)\rightarrow \frac{1}{3}\delta^{aa'}\bar q(x)\Gamma_i q(0)$, the Eq. (\ref{QCDES}) takes the form
 \begin{eqnarray}
\label{QCDES2}
\Pi _{\mu \nu-Nonpert }^{\mathrm{QCD}-Mol}(p,q)&=&-i \int d^{4}xe^{ip\cdot x} \Bigg\{  \mathrm{Tr}%
\big[ \gamma _{\mu }\Gamma_i\gamma _{\nu
}  
%\nonumber\\
%& \times 
S_{b}^{b^{\prime }b}(-x)\big]
\mathrm{Tr}\big[ \gamma _{5}{S}_{c}^{aa^{\prime
}}(x)\gamma _{5}S_{q}^{a^{\prime  }a}(-x)\big] \delta^{bb^{\prime }}
\nonumber\\
&&+
\mathrm{Tr}%
\big[ \gamma _{\mu }{S}_{q}^{bb^{\prime }}(x)\gamma _{\nu
}  
%\nonumber\\
%& \times 
S_{b}^{b^{\prime }b}(-x)\big]
\mathrm{Tr}\big[ \gamma _{5}{S}_{c}^{aa^{\prime
}}(x)\gamma _{5}\Gamma_i \big]\delta^{a^{\prime  }a}
\Bigg\} \frac{1}{12} \langle \gamma(q) |\bar q(x)\Gamma_i q(0)|0\rangle . 
 \end{eqnarray}

 In addition to the above computations, when a light quark interacts with a photon nonperturbatively, a gluon can also be released from one of the remaining three quark propagators. 
The expression obtained after performing these calculations is as follows:
\begin{eqnarray}
\label{QCDES4}
\Pi _{\mu \nu-Nonpert }^{\mathrm{QCD}-Mol}(p,q)&=&-i\int d^{4}xe^{ip\cdot x}  \Bigg\{  \mathrm{Tr}%
\big[ \gamma _{\mu }\Gamma_i\gamma _{\nu
}  
%\nonumber\\
%& \times 
S_{b}^{b^{\prime }b}(-x)\big]
\mathrm{Tr}\big[ \gamma _{5}{S}_{c}^{aa^{\prime
}}(x)\gamma _{5}S_{q}^{a^{\prime  }a}(-x)\big]
%%%%%%%%%%%%%%%%%%%%%55
\Big[\Big(\delta^{bb'}\delta^{b'b}-\frac{1}{3}\delta^{bb'}\delta^{b'b}\Big)\nonumber\\
&&+\Big(\delta^{ba}\delta^{b'a'}-\frac{1}{3}\delta^{bb'}\delta^{aa'}\Big)
+\Big(\delta^{ba'}\delta^{b'a}-\frac{1}{3}\delta^{bb'}\delta^{a'a}\Big)\Big]
%%%%%%%%%%%%%%%%%%%%%%%%%%%%%%%%
%%%%%%%%%%%%%%%%%%%%%%%%%%%%%%%%%%
\nonumber\\
&&+\mathrm{Tr}%
\big[ \gamma _{\mu }{S}_{q}^{bb^{\prime }}(x)\gamma _{\nu
}  
%\nonumber\\
%& \times 
S_{b}^{b^{\prime }b}(-x)\big]
\mathrm{Tr}\big[ \gamma _{5}{S}_{c}^{aa^{\prime
}}(x)\gamma _{5}\Gamma_i \big]
%%%%%%%%%%%%%%%%%%%%%%%%%%%%%%%%%%%%5
\Big[ \Big(\delta^{ba'} \delta^{b'a}-\frac{1}{3} \delta^{bb'}\delta^{a'a}\Big)\nonumber\\
&&+\Big(\delta^{b'a'} \delta^{ba}-\frac{1}{3} \delta^{b'b}\delta^{a'a}\Big)
+\Big(\delta^{aa'} \delta^{a'a}-\frac{1}{3} \delta^{aa'}\delta^{a'a}\Big)
%%%%%%%%%%%%%%%%%%%%5
\Bigg\}
\nonumber\\
&& \times \frac{1}{32} \langle \gamma(q) |\bar q(x)\Gamma_i G_{\mu\nu}(vx) q(0)|0\rangle ,
 \end{eqnarray}
where we used
\begin{align}
\label{QCDES5}
 \bar q^a(x)\Gamma_i G_{\mu\nu}^{bb'}(vx) q^{a'}(0)\rightarrow \frac{1}{8}\Big(\delta^{ab}\delta^{a'b'}
 -\frac{1}{3}\delta^{aa'}\delta^{bb'}\Big)\bar q(x)\Gamma_i G_{\mu\nu}(vx) q(0).
\end{align}

  We observe that matrix elements of the form  $\langle \gamma(q) | \bar{q}(x) \Gamma_i q(0) | 0\rangle$ and $\langle \gamma(q)\vel \bar{q}(x) \Gamma_i G_{\mu\nu}q(0) \ver 0\rangle$ are required.
These matrix elements are parameterized concerning the photon distribution amplitudes (DAs) and explicit expressions of these DAs are given in Ref.~\cite{Ball:2002ps}. In addition to these matrix elements, non-local operators such as four quarks ($\bar qq \bar q q$) and two gluons ($\bar q G G q$) can also be seen in the calculations. However it is known that the effects of such nonlocal operators are small \cite{Balitsky:1987bk,Braun:1989iv}, and hence we will neglect them.
Using Eqs. (\ref{new:QCDSide1}), (\ref{QCDES2}) and (\ref{QCDES4}), the calculations of the QCD side of the correlator of the analysis are obtained.

\section*{Appendix B: Explicit forms of the  \texorpdfstring{$\Delta_i^{QCD}(M^2,s_0)$}{} functions } 
%%%
In this appendix, we give the explicit expression for the  $\Delta_1^{QCD}(M^2,s_0)$ and  $\Delta_2^{QCD}(M^2,s_0)$ functions: 
\begin{align}
\Delta_1^{QCD}(M^2,s_0)&=
-\frac {e_b} {15728640 \pi^5} \Bigg[ 1440  m_b P_ 2 \pi^2 \Big (I[0, 4, 1, 0] - 3 I[0, 4, 1, 1] + 
    3 I[0, 4, 1, 2] - I[0, 4, 1, 3] - 
    3 (I[0, 4, 2, 0]  
    \nonumber\\
%           \end{align}
%          \begin{align}
    %%%%%%%%%%%%%%%%%%%%%%%%%%%%%%%%
   %     \end{align}
   %     \begin{align}
        &- 2 I[0, 4, 2, 1]+ I[0, 4, 2, 2] - 
       I[0, 4, 3, 0] + I[0, 4, 3, 1]) - I[0, 4, 4, 0]\Big) + 
 480 m_ 0^2 m_c P_ 2 \pi^2 \Big (I[0, 3, 2, 0] 
    \nonumber\\
    &- 2 I[0, 3, 2, 1] + 
    I[0, 3, 2, 2] - 2 I[0, 3, 3, 0] + 2 I[0, 3, 3, 1] + 
    I[0, 3, 4, 0] + 6 I[1, 2, 2, 1] - 3 I[1, 2, 2, 2]
    \nonumber\\
        & - 
    6 I[1, 2, 3, 1]\Big) + 
 10 (- P_ 1 + 48 m_c P_ 2 \pi^2) \Big (I[0, 4, 2, 0] - 
    3 I[0, 4, 2, 1] + 3 I[0, 4, 2, 2] - I[0, 4, 2, 3] 
    \nonumber\\
            &- 
    2 I[0, 4, 3, 0] + 4 I[0, 4, 3, 1] - 2 I[0, 4, 3, 2] + 
    I[0, 4, 4, 0] - I[0, 4, 4, 1] + 4 I[1, 3, 2, 1] - 
    8 I[1, 3, 2, 2] 
    \nonumber\\
&+ 4 I[1, 3, 2, 3] - 8 I[1, 3, 3, 1] + 
    8 I[1, 3, 3, 2] + 4 I[1, 3, 4, 1]\Big) - 
 9 \Big (2 I[0, 6, 2, 1] - 7 I[0, 6, 2, 2] 
    \nonumber\\
            & - 
    5 I[0, 6, 2, 4] + I[0, 6, 2, 5] - 6 I[0, 6, 3, 1]+ 
    15 I[0, 6, 3, 2] - 12 I[0, 6, 3, 3] + 3 I[0, 6, 3, 4] + 
    6 I[0, 6, 4, 1] 
    \nonumber\\
&- 9 I[0, 6, 4, 2] + 3 I[0, 6, 4, 3] - 
    2 I[0, 6, 5, 1] + I[0, 6, 5, 2] + 6 I[1, 5, 2, 2]- 
    18 I[1, 5, 2, 3] + 18 I[1, 5, 2, 4]
    \nonumber\\
& - I[1, 5, 2, 5] - 
    18 I[1, 5, 3, 2] - 18 I[1, 5, 3, 3] + 18 I[1, 5, 3, 4] - 
    18 I[1, 5, 4, 2] + 18 I[1, 5, 4, 3] \nonumber\\
    &- I[1, 5, 5, 2]\Big)\Bigg]\nonumber\\
&+\frac {5 e_c P_ 1} {56623104 \pi^5}  \Bigg[12 I[0, 4, 2, 1]-4 I[0, 4, 2, 0] 
     - 12 I[0, 4, 2, 2] + 
    4 I[0, 4, 2, 3] + 5 I[0, 4, 3, 0] - 10 I[0, 4, 3, 1] \nonumber\\
    &+ 5 I[0, 4, 3, 2] +2 I[0, 4, 4, 
      0] -2 I[0, 4, 4, 1] - 3  I[0, 4, 5, 0] - 
    32  m_b \Bigg (4 P_ 2 \pi^2\Big (I[0, 2, 1, 0] - 2 I[0, 2, 1, 1]  \nonumber\\
           &+
           I[0, 2, 1, 2]- 2 I[0, 2, 2, 0] + 2 I[0, 2, 2, 1] + 
          I[0, 2, 3, 0] - 2 I[1, 1, 1, 0] - 4 I[1, 1, 1, 1] + 
          2 I[1, 1, 1, 2]\nonumber\\
                            & - 4 I[1, 1, 2, 0]+ 4 I[1, 1, 2, 1] + 
          2 I[1, 1, 3, 0]\Big) \Bigg) - 16 I[1, 3, 2, 1] - 
    32 I[1, 3, 2, 2] + 16 I[1, 3, 2, 3] \nonumber\\
        & - 8 I[1, 3, 3, 1] + 
    20 I[1, 3, 3, 2]- 8 I[1, 3, 4, 1]\Bigg]\nonumber\\
               %        \end{align}
   %     \begin{align}
                           &+\frac {e_q} {5242880 \pi^5} \Bigg[  5  P_ 1 \Big (I[0, 4, 3, 0] - 2 I[0, 4, 3, 1] + I[0, 4, 3, 2] - 
    2 I[0, 4, 4, 0] + 2 I[0, 4, 4, 1] + I[0, 4, 5, 0] 
    \nonumber\\
&+ 
    8 I[1, 3, 3, 1] - 4 I[1, 3, 3, 2] - 8 I[1, 3, 4, 1]\Big) - 
 240 m_c P_ 2 \pi^2 \Big (I[0, 4, 3, 0] - 2 I[0, 4, 3, 1] + 
    I[0, 4, 3, 2] 
    \nonumber\\
& - 2 I[0, 4, 4, 0] + 2 I[0, 4, 4, 1]+ 
    I[0, 4, 5, 0] + 8 I[1, 3, 3, 1] - 4 I[1, 3, 3, 2] - 
    8 I[1, 3, 4, 1]\Big) - 
 9 \Big (I[0, 6, 3, 0] 
    \nonumber\\
&- 4 I[0, 6, 3, 1] + 6 I[0, 6, 3, 2] - 
    4 I[0, 6, 3, 3]+ I[0, 6, 3, 4] - 3 I[0, 6, 4, 0] + 
    9 I[0, 6, 4, 1] - 9 I[0, 6, 4, 2]  
    \nonumber\\
&+ 3 I[0, 6, 4, 3] + 
    3 I[0, 6, 5, 0] - 6 I[0, 6, 5, 1] + 3 I[0, 6, 5, 2]- 
    I[0, 6, 6, 0] + I[0, 6, 6, 1] + 6 I[1, 5, 3, 1] 
    \nonumber\\
&- 
    18 I[1, 5, 3, 2] + 18 I[1, 5, 3, 3] - 6 I[1, 5, 3, 4] - 
    18 I[1, 5, 4, 1] + 36 I[1, 5, 4, 2]- 18 I[1, 5, 4, 3] \nonumber\\
    &+ 
    18 I[1, 5, 5, 1] - 18 I[1, 5, 5, 2] - 6 I[1, 5, 6, 1]\Big)\Bigg]\nonumber\\
 %                 \end{align}
  %            \begin{align}
                %%%%%%%%%%%%%%%%%%%%%%%%%%%%%
                %%%%%%%%%%%%%%%%%%%%%%%%%%%%%%%
                &+m_0^2\Big( 17 e_q I_4[\mathcal S]\, I[0, 5, 3, 0]+ 2 e_q\, I_4[\mathcal T_1] I[0, 5, 4, 0]+476\, e_q\, f_{3\gamma}\, \pi^2\, I_2[\mathcal V]\, I[0, 4, 4, 0]
\Big) + \chi\,m_0^2\Big(238\, e_q \nonumber\\
& \times I_1[\mathcal V]\, I[0, 6, 4, 0]
+ 13 e_q I_3[\mathcal S]\, I[0, 6, 3, 0]+ 8 e_q\, I_3[\mathcal T_1] I[0, 6, 2, 0]\Big),
\label{app}
\end{align}
\begin{align}
\Delta^{QCD}_2&=\frac {e_b} {11796480 \pi^5} \Bigg[
   960\, m_ 0^2 \, m_c \,P_ 2 \pi^2 \Big (I[0, 3, 2, 0] - 2 I[0, 3, 2, 1] + 
       I[0, 3, 2, 2] - 2 I[0, 3, 3, 0] + 2 I[0, 3, 3, 1] \nonumber\\
       &+ 
       I[0, 3, 4, 0] + 6 I[1, 2, 2, 1] - 3 I[1, 2, 2, 2] - 
       6 I[1, 2, 3, 1]\Big) + 
    5  \Bigg (576 m_b P_ 2 \pi^2\Big (I[0, 4, 1, 0] - 
          3 I[0, 4, 1, 1] \nonumber\\
 %              \end{align}
%\begin{align}
    %%%%%%%%%%%%%%%%%%%%%%%%%%%%%%%%
          &+ 3 I[0, 4, 1, 2] - I[0, 4, 1, 3] - 
          3 (I[0, 4, 2, 0] - 2 I[0, 4, 2, 1] + I[0, 4, 2, 2] - 
             I[0, 4, 3, 0] + I[0, 4, 3, 1]) - 
          I[0, 4, 4, 0]\Big)\nonumber\\
          &+ (P_ 1 - 
           192  m_c P_ 2 \pi^2) \Big (I[0, 4, 2, 0] - 
           3 I[0, 4, 2, 1] + 3 I[0, 4, 2, 2] - I[0, 4, 2, 3] - 
           2 I[0, 4, 3, 0] + 4 I[0, 4, 3, 1] \nonumber\\
           &- 2 I[0, 4, 3, 2] + 
           I[0, 4, 4, 0] - I[0, 4, 4, 1] + 
           4 \big (I[1, 3, 2, 1] - 2 I[1, 3, 2, 2] + I[1, 3, 2, 3] - 
               2 I[1, 3, 3, 1] + 2 I[1, 3, 3, 2] \nonumber\\
         %       \end{align}
         % \begin{align}
           &+ 
               I[1, 3, 4, 1]\big)\Big)\Bigg)       
                - 
    18 \Big (2 I[0, 6, 2, 1]- 7 I[0, 6, 2, 2] + 9 I[0, 6, 2, 3] - 
        5 I[0, 6, 2, 4] + I[0, 6, 2, 5] - 6 I[0, 6, 3, 1]   \nonumber\\
        &+ 
        15 I[0, 6, 3, 2]- 12 I[0, 6, 3, 3]+ 3 I[0, 6, 3, 4] + 
        6 I[0, 6, 4, 1] - 9 I[0, 6, 4, 2] + 3 I[0, 6, 4, 3] - 
        2 I[0, 6, 5, 1]   \nonumber\\
        &+ I[0, 6, 5, 2]+ 
        6 \big (I[1, 5, 2, 2]- 3 I[1, 5, 2, 3] + 3 I[1, 5, 2, 4] - 
            I[1, 5, 2, 5] - 
            3 (I[1, 5, 3, 2] - 2 I[1, 5, 3, 3]  \nonumber\\
               &+ I[1, 5, 3, 4] - 
               I[1, 5, 4, 2]+ I[1, 5, 4, 3])- 
            I[1, 5, 5, 2]\big)\Big)\Bigg]\nonumber\\
            %%%%%%%%%%%%%%%%%%%%%%%%%%%%%%%%%%%%%%%%%%%%%%%
           & +\frac {5 e_c P_ 1} {56623104 \pi^5}  \Bigg[12 I[0, 4, 2, 1]-4 I[0, 4, 2, 0] 
     - 12 I[0, 4, 2, 2] + 
    4 I[0, 4, 2, 3] + 5 I[0, 4, 3, 0] - 10 I[0, 4, 3, 1] \nonumber\\
    &+ 5 I[0, 4, 3, 2] +2 I[0, 4, 4, 
      0] -2 I[0, 4, 4, 1] - 3  I[0, 4, 5, 0] - 
    64  m_b \Bigg ( P_ 2 \pi^2\Big (I[0, 2, 1, 0] - 2 I[0, 2, 1, 1]  \nonumber\\
           &+
           I[0, 2, 1, 2]- 2 I[0, 2, 2, 0] + 2 I[0, 2, 2, 1] + 
          I[0, 2, 3, 0] - 2 I[1, 1, 1, 0] - 4 I[1, 1, 1, 1] + 
          2 I[1, 1, 1, 2] - 4 I[1, 1, 2, 0]\nonumber\\
          &+ 4 I[1, 1, 2, 1] + 
          2 I[1, 1, 3, 0]\Big) \Bigg) - 16 I[1, 3, 2, 1] - 
    16 I[1, 3, 2, 2] + 32 I[1, 3, 2, 3] - 8 I[1, 3, 3, 1] + 
    24 I[1, 3, 3, 2] \nonumber\\
    &- 16 I[1, 3, 4, 1]\Bigg]\nonumber\\
    %%%%%%%%%%%%%%%%%%%%%%%%%%%%%%%%%%%%%%%%%%%%%%%%%%%%%%%%%5
                    &
            +\frac {e_q} {141557760 \pi^5} \Bigg[-800 m_b P_ 1 \Bigg (4 P_ 2 \
\pi^2 \Big (I[0, 2, 1, 0] - 2 I[0, 2, 1, 1] + I[0, 2, 1, 2] - 
          2 I[0, 2, 2, 0] + 2 I[0, 2, 2, 1] \nonumber\\
              &+ I[0, 2, 3, 0] - 
          2 \big (I[1, 1, 1, 0] - 2 I[1, 1, 1, 1] + I[1, 1, 1, 2] - 
              2 I[1, 1, 2, 0] + 2 I[1, 1, 2, 1] + 
              I[1, 1, 3, 0]\big)\Big)\nonumber%\\
                                %   \end{align}
             % \begin{align}
            \Bigg)\nonumber\\
            %%%%%%%%%%%%%%%%%%%%%%%%%%%%%%%%%%%%%%%%%%%
        %        \begin{align}
      &- 
    5 P1 \Big (20 I[0, 4, 2, 0] - 60 I[0, 4, 2, 1] + 
       60 I[0, 4, 2, 2] - 20 I[0, 4, 2, 3] - 7 I[0, 4, 3, 0] + 
       14 I[0, 4, 3, 1] - 7 I[0, 4, 3, 2] \nonumber\\
       &- 46 I[0, 4, 4, 0] + 
       46 I[0, 4, 4, 1] + 33 I[0, 4, 5, 0] + 
       4 \big (20 I[1, 3, 2, 1] - 40 I[1, 3, 2, 2] + 
           20 I[1, 3, 2, 3] + 26 I[1, 3, 3, 1] \nonumber\\
                     &+ 7 I[1, 3, 3, 2] - 
           46 I[1, 3, 4, 1]\big)\Big) - 
    216 \Bigg (4 m_c \Big (20 P_ 2 \pi^2 \big (I[0, 4, 3, 0] - 
              2 I[0, 4, 3, 1] + I[0, 4, 3, 2] - 2 I[0, 4, 4, 0] \nonumber\\
              &+ 
              2 I[0, 4, 4, 1] + I[0, 4, 5, 0] + 8 I[1, 3, 3, 1] - 
              4 I[1, 3, 3, 2] - 8 I[1, 3, 4, 1]\big) \Big) + 
        3 \Big (I[0, 6, 3, 0] - 4 I[0, 6, 3, 1]   \nonumber\\
            &+ 6 I[0, 6, 3, 2]+ I[0, 6, 3, 4] - 3 I[0, 6, 4, 0] + 
            9 I[0, 6, 4, 1] - 9 I[0, 6, 4, 2] + 3 I[0, 6, 4, 3] + 
            3 I[0, 6, 5, 0] \nonumber\\
            &- 6 I[0, 6, 5, 1] + 3 I[0, 6, 5, 2]- 
            I[0, 6, 6, 0] + I[0, 6, 6, 1] + 
            6 \big (I[1, 5, 3, 1] - 3 I[1, 5, 3, 2] + 
                3 I[1, 5, 3, 3]  \nonumber\\
                &- I[1, 5, 3, 4] - 
                3 (I[1, 5, 4, 1] - 2 I[1, 5, 4, 2]+ I[1, 5, 4, 3] - 
                   I[1, 5, 5, 1] + I[1, 5, 5, 2]) - 
                I[1, 5, 6, 1]\Big)\Big)\Bigg)\Bigg]\nonumber\\
               % \end{align}
                %\begin{align}
                %%%%%%%%%%%%%%%%%%%%%%%%%%%%%
                %%%%%%%%%%%%%%%%%%%%%%%%%%%%%%%
                &+32 m_0^2\Bigg[\Big( 26 e_q I_4[\mathcal S]\, I[0, 5, 3, 0]+ 12 e_q\, I_4[\mathcal T_1] I[0, 5, 4, 0]+238\, e_q\, f_{3\gamma}\, \pi^2\, I_2[\mathcal V]\, I[0, 4, 4, 0]
\Big) + \chi\,\Big(476\, e_q\,I_1[\mathcal V] \nonumber\\
& \times \, I[0, 6, 5, 0]
+ 26 e_q I_3[\mathcal S]\, I[0, 6, 4, 0]+ 12 e_q\, I_3[\mathcal T_1] I[0, 6, 3, 0]\Big)\Bigg],
\label{app1}
\end{align}
where $P_1 =\langle g_s^2 G^2\rangle$, $P_2 =\langle \bar q q \rangle$ are gluon and u/d-quark condensates, respectively. We should also mention that, in Eqs. (\ref{app}) and (\ref{app1}), for the sake of brevity we have only presented the terms that give substantial contributions to the numerical values of the observables under investigation and ignored to present many higher dimensional operators though they have been considered in the numerical calculations.

The functions~$I[n,m,l,k]$, $I_1[\mathcal{F}]$,~$I_2[\mathcal{F}]$,~$I_3[\mathcal{F}]$ and~$I_4[\mathcal{F}]$ are
defined as:
\begin{align}
 I[n,m,l,k]&= \int_{(m_c+m_b)^2}^{s_0} ds \int_{0}^1 dt \int_{0}^1 dw~ e^{-s/M^2}~
 s^n\,(s-(m_c+m_b)^2)^m\,t^l\,w^k,\nonumber\\
 %   \end{align}
 %\begin{align}
 I_1[\mathcal{F}]&=\int D_{\alpha_i} \int_0^1 dv~ \mathcal{F}(\alpha_{\bar q},\alpha_q,\alpha_g)
 \delta'(\alpha_ q +\bar v \alpha_g-u_0),\nonumber\\
%  \end{align}
% \begin{align}
  I_2[\mathcal{F}]&=\int D_{\alpha_i} \int_0^1 dv~ \mathcal{F}(\alpha_{\bar q},\alpha_q,\alpha_g)
 \delta'(\alpha_{\bar q}+ v \alpha_g-u_0),\nonumber\\
 %  \end{align}
% \begin{align}
   I_3[\mathcal{F}]&=\int D_{\alpha_i} \int_0^1 dv~ \mathcal{F}(\alpha_{\bar q},\alpha_q,\alpha_g)
 \delta(\alpha_ q +\bar v \alpha_g-u_0),\nonumber\\
 %  \end{align}
% \begin{align}
   I_4[\mathcal{F}]&=\int D_{\alpha_i} \int_0^1 dv~ \mathcal{F}(\alpha_{\bar q},\alpha_q,\alpha_g)
 \delta(\alpha_{\bar q}+ v \alpha_g-u_0),\nonumber
 %   I_5[\mathcal{F}]&=\int_0^1 du~ \mathcal{F}(u)\delta'(u-u_0),\nonumber\\
 %  \end{align}
% \begin{align}
% I_6[\mathcal{F}]&=\int_0^1 du~ \mathcal{F}(u),\nonumber
 \end{align}
 where $\mathcal{F}$ stands for the corresponding photon DAs.
 
 \end{widetext}
 
% \textbf{Data Availability Statement:} This manuscript has no associated data or the data will not be deposited. [Authors’ comment: This is a theoretical research work, so no additional data are associated with this work.]

\bibliography{ZcbMM}

\end{document}